# Demonstration of a polariton step potential by local variation of light-matter coupling in a van-der-Waals heterostructure

C. RUPPRECHT,[1] M. KLAAS,[1] H. KNOPF,[2,3,4] T. TANIGUCHI,[5] K. WATANABE,[5] Y. QIN,[6] S. TONGAY,[6] S. SCHRÖDER,[3] F. EILENBERGER,[2,3,4] S. HÖFLING[1,7] AND C. SCHNEIDER[1,*]

[1] *Technische Physik and Wilhelm-Conrad-Röntgen-Research Center for Complex Material Systems, Universität Würzburg, D-97074 Würzburg, Am Hubland, Germany.*
[2] *Institute of Applied Physics, Abbe Center of Photonics, Friedrich Schiller University, Albert-Einstein-Straße 15, 07745 Jena, Germany*
[3] *Fraunhofer-Institute for Applied Optics and Precision Engineering IOF, Albert-Einstein-Straße 7, 07745 Jena*
[4] *Max Planck School of Photonics, Albert-Einstein-Straße 7, 07745 Jena, Germany*
[5] *National Institute for Materials Science, Tsukuba, Ibaraki 305-0044, Japan*
[6] *Arizona State University, Tempe, AZ 85287, USA*
[7] *SUPA, School of Physics and Astronomy, University of St. Andrews, St. Andrews KY 16 9SS, United Kingdom*
*\*christian.schneider@physik.uni-wuerzburg.de*

**Abstract:** The large oscillator strength of excitons in transition metal dichalcogenide layers facilitates the formation of exciton-polariton resonances for monolayers and van-der-Waals heterostructures embedded in optical microcavities. Here, we show, that locally changing the number of layers in a $WSe_2$/hBN/$WSe_2$ van-der-Waals heterostructure embedded in a monolithic, high-quality-factor cavity gives rise to a local variation of the coupling strength. This effect yields a polaritonic stair case potential, which we demonstrate at room temperature. Our result paves the way towards engineering local polaritonic potentials at length scales down to atomically sharp interfaces, based on purely modifying its real part contribution via the coherent light-matter coupling strength g.

## 1. Introduction

The regime of strong coupling between confined excitons and microcavity photons was first observed in 1992 by Weisbuch et al. in a semiconductor microcavity with embedded GaAs quantum wells [1]. Light-matter hybridization allows to create quasi-particles with precisely tailored properties, such as well-defined effective masses and interaction constants [2]. Exciton-polaritons have become a testbed to study collective excitations and coherent phenomena such as Bose-Einstein condensation in solid state systems at elevated temperatures [3-5], topological excitations [6,7], and superfluidity [8]. Furthermore, polariton condensates have been recently proposed as a platform for classical- and quantum simulators, in structures with well-defined polariton potentials [9-11]. The latter have mostly been engineered by either trapping the photonic part of the polariton wave function, or by locally changing the detuning of the exciton [12]. Most recently, polariton trapping and potential landscape engineering techniques have been applied to room-temperature exciton-polariton condensates [13,14] as well as structures operated under electrical current injection [15]. Here, we use polaritons composed of excitonic emitters in atomically thin transition metal dichalcogenide (TMDC) layers [16-20] and van-der-Waals heterostructures [17] embedded in a photonic cavity. We utilize the dependency of the polaritonic potential to a local change of thickness of the van-der-Waals heterostructure, to locally change the light-matter coupling strength g and the photonic cavity resonance. This is facilitated by embedding a van-der-Waals heterostructure composed of two $WSe_2$ monolayers, separated by a thin (5 nm) layer of hBN in a dielectric cavity of high quality factor. We utilize highly spatially and momentum resolved spectroscopy to demonstrate a polariton step-like potential at the interface between the single monolayer region and the double monolayer region, where two monolayers are separated with h-BN.

**2. Sample structure and basic characterization**

Fig. 1a depicts a graphic illustration of our microcavity device: It consists of a $SiO_2/TiO_2$ distributed Bragg reflector (DBR) (10 pairs) as the bottom mirror, with a central Bragg wavelength at 750 nm, an embedded van-der-Waals heterostructure (orange), as well as a $SiO_2/TiO_2$ top DBR. We utilize the Scotch tape method combined with a dry transfer method to implement a layer sequence of $WSe_2$/h-BN/$WSe_2$ (Fig. 1a,b) on the bottom DBR, i.e. two monolayers $WSe_2$ separated by a h-bN multilayer. First, we micromechanically cleave a high quality, CVD-grown $WSe_2$ crystal, as well as the BN crystal via sticky tape. Next, we bring the tape in contact with a PolyDimethyl-Siloxane (PDMS) stamp. Atomically thin $WSe_2$ crystals, as well as suitable h-BN flakes are then identified by their colour contrast in a high resolution optical microscope. We then use the standart dry stamping technique to transfer the flakes onto the bottom DBR by mounting them upside down on a home-built Micromanipulator, attached to a manual xyz-stage in the microscope. The combination of a cross on the microscope objective and this stage made it possible to transfer the layers with a precision better than $\pm 2\mu m$ on the bottom DBR. We note, that the top TMDC layer is slightly smaller than the bottom layer, yielding a well-defined interface region in the heterostack. While the bottom DBR was purchased from the company Laseroptics GmbH with a central wavelength at 750 nm, the top DBR composed of 9 repetitions of SiO2/TiO2 is deposited on top of the structure utilizing a gentle low temperature technique [21] with 129 nm/ 83 nm layer thicknesses. This results in a physical cavity length of 234 nm, whereas the first 105 nm SiO2 layer were coated with lower density to protect the TMDC layers from cracking [21]. The quality factor of the sample, which we measure via white-light reflection on the empty cavity regions next to the van-der-Waals stack was determined to be $Q = 4760 \pm 370$ and associated with a linewidth of $(0.360 \pm 0.010)$ meV. (Fig 1c,d).

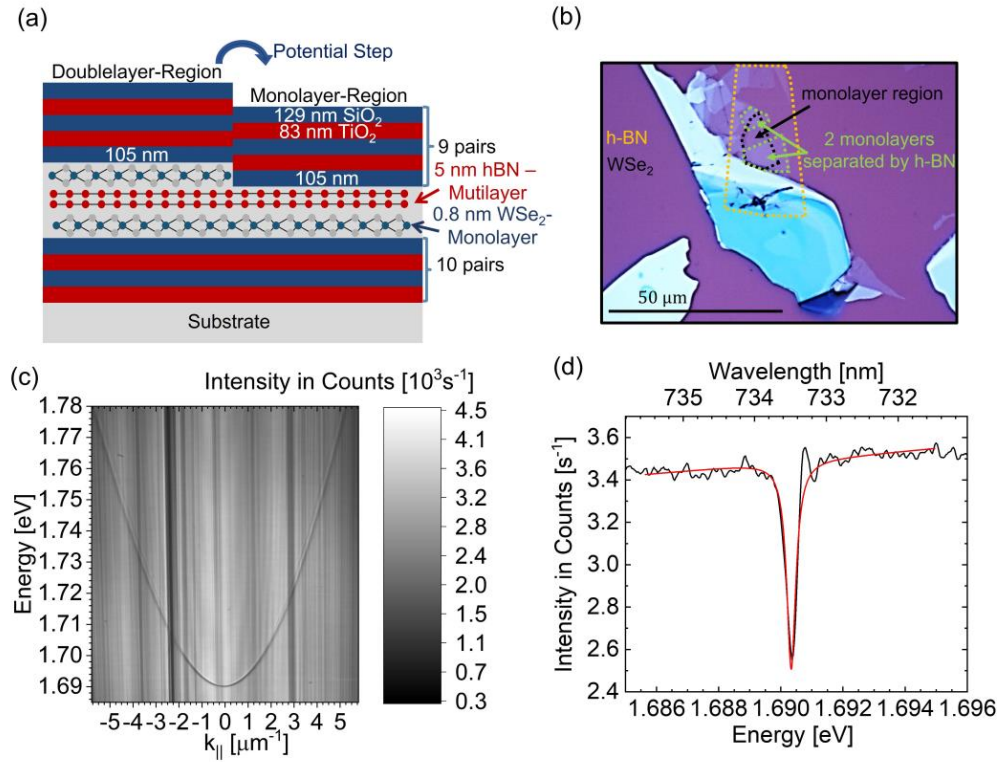

Fig. 1. a) Schematic of the fully grown microcavity structure with embedded $WSe_2$/h-BN/ $WSe_2$ heterostructure (orange) b) A microscopic top-view of the $WSe_2$/h-BN/ $WSe_2$ structure, while the lower/ upper monolayer is indicated with the black/ green dashed line c) White light reflection spectrum showing the empty cavity resonance in momentum space d) The line spectrum of the empty cavity resonance at $k = 0\mu m^{-1}$ fitted by a Lorentzian revealing a $Q = 4760 \pm 370$.

Prior to capping the structure with the top DBR, we probed the optical response of the single monolayer, as well as the heterostructures composed of the two monolayers separated by h-BN region by optical absorption spectroscopy utilizing a broad-band thermal light-source. All photoluminescence measurements were performed with a 532 nm CW laser. In the single monolayer region, the room-temperature absorption contrast spectrum is characterized by a single absorption signal at an energy of 1.675 eV and linewidth of $(59.0 \pm 1.0)$ meV (Fig. 2a). This signal shifts towards 1.653 eV in the heterostructure area, the linewidth broadens to $(82.6 \pm 0.64)$ meV (Fig. 2b). Both spectra were normalized to a reference signal, measured next to the heterostructure. We believe that the presence of the second $WSe_2$ layer, separated by ~ 5 nm hBN, modifies the excitonic energy via a modification of dielectric screening, yielding the observed spectral shifts. We note, that red-shifts of similar magnitude have been observed in graphene and graphite capped $WS_2$ samples [22] or in TMDC-TMDC heterostructures [23].

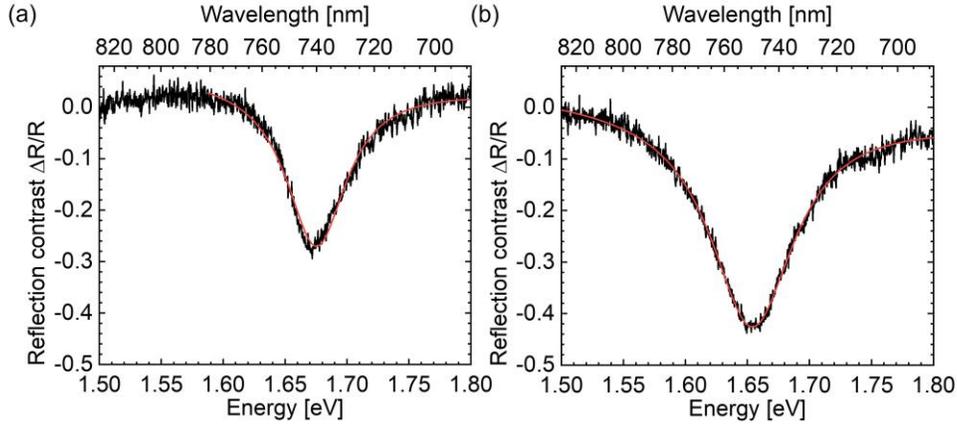

Fig. 2. Absorption spectra of a monolayer a) and heterostructure b) with linewidths of $(59.0 \pm 1.0)$ meV and $(82.6 \pm 0.64)$ meV. These were extracted from fitted Lorentzians (red). Both spectra were normalized on a reference signal that was recorded next to the heterostructure.

## 3. Results and discussion

Next, we study the photoluminescence from our device via momentum-resolved spectroscopy at room temperature. We used an optical setup in which both spatially (near-field) and momentum-space (far-field) resolved spectroscopy and imaging are accessible. PL is collected through a 0.65 NA microscope objective, and directed into an imaging spectrometer with up to 1200 groves/mm grating via a set of relay lenses, projecting the proper projection plane onto the monochromator's entrance slit. The system's angular resolution is ~ 0.03 μm-1 (~ 0.2°) and its spectral resolution is up to ~ 0.050 meV with a Peltier–cooled Si-CCD as detector.

Fig 3a) and b) show the momentum resolved photoluminescence from the device, recorded at positions where only the single monolayer (a) or the full heterostructure (b) contribute to the coupling. In the single monolayer area, photoluminescence emerges at an energy of 1.643 eV, and acquires a curved dispersion relation. For the area of the complete van-der-Waals stack, a dispersive signal is present which is red-shifted by 31 meV. We further observe emission signals from the empty cavity dispersion at ~1.68 eV (at $k = 0$) as well as 1.69 eV, which stem from regions outside of the monolayer. The trace of both signals can be modeled in a straightforward manner utilizing the standard coupled oscillator model describing the normal mode coupling of exciton and photon

$$\begin{bmatrix} E_{ex} & V/2 \\ V/2 & E_{ph} \end{bmatrix} \begin{bmatrix} \alpha \\ \beta \end{bmatrix} = E \begin{bmatrix} \alpha \\ \beta \end{bmatrix}, \quad (1)$$

where $E_{ph}$ and $E_{ex}$ are photon and exciton energies, respectively, and V the exciton-photon coupling strength. The eigenvectors yield the Hopfield coefficients for the exciton and photon fractions of the polariton states.

The result of this modelling is shown in Fig. 3c, where the corresponding line spectra for the monolayer polariton and the heterostructure polariton were extracted from Fig. 3a and b. For the fitting procedure the initial values for the exciton were set to the minima in the branches. Dashed red and blue lines show exciton and cavity resonances of monolayer and heterostructure polariton, respectively. The green mode corresponds to the empty cavity resonance with h-BN.

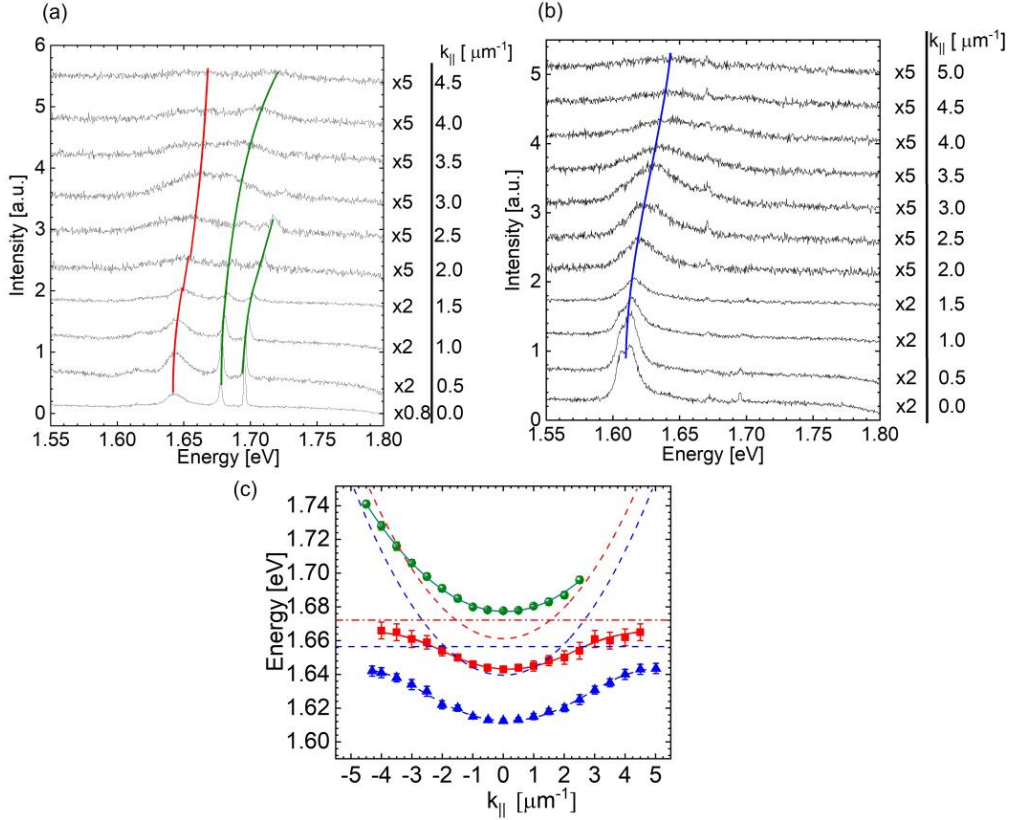

Fig. 3. a) Line spectra showing the monolayer polariton (red guideline). The low energetic green line indicates an empty cavity + h-BN and the other an empty cavity resonance. Extracted from Fig. 5b). b) Line spectra showing the heterostructure polartion (blue guideline). Extracted from Fig.5a). c) The extracted peak positons with uncertainties from previous graphs and the coupled oscillator fit. Red and blue data corresponds to monolayer and heterostructure polariton

absorption dips and the cavity effective mass as well as the energetic minimum was set equal to the h-BN mode. In principle, the empty cavity should not be visible in the strong coupling regime, but we assume that PL from edge regions or from smaller monolayer pieces in close proximity weakly couple to the empty cavity mode. From this fit, we extract an increase of the coupling strength from $(46.2 \pm 6.3)$ meV to $(69.5 \pm 7.4)$ meV, which is an enhancement of V by a factor of 1.5 $(\pm 0.13)$. This agrees well with the expected scaling of the coupling strength with the square root of the number of oscillators (monolayers n) $V \propto n^{1/2}$ [2,17]. Additionally, the ratio between oscillator strengths in Fig. 2 extracted from the dip areas leads to an increase of the coupling strength of $1.565 \pm 0.032$, what is in good agreement with our measurements. As can be seen in Fig. 3a),c), the mode energy decreases as going from the empty cavity (Fig. 3a), ,high energetic green line at $1.6901$ eV for $k = 0\,\mu m^{-1}$) to cavity + h- BN (Fig. 3a), low energetic green line and Fig. 3c) green line at $1.6773$ eV for $k = 0\,\mu m^{-1}$) to WSe$_2$/h-BN (Fig. 3c), red dashed line at $1.6612$ eV for $k = 0\,\mu m^{-1}$) to WSe$_2$/h-BN/WSe$_2$ (Fig. 3c), blue dashed line at $1.6396$ eV for $k = 0\,\mu m^{-1}$). From AFM measurements of reference samples, we know, that our exfoliated h-BN is $\sim 5 - 10$ nm thick. When going from an empty cavity to cavity + hBN, the experimentally observed spectral shift of the cavity mode of $12.8$ meV corresponds with a 5nm thick hBN layer. When going from cavity + hBN to WSe$_2$/h-BN ($16.1$ meV shift) and WSe$_2$/h-BN/WSe$_2$ ($37.7$ meV shift), the significant optical thickness of WSe$_2$ arises due to a high refractive index of around 4.2 at 750 nm (and room temperature). Indeed, we have checked the magnitude of this spectral shift, by simulating a microcavity with "passive" TMDC material, where we ignored the imaginary part of the refractive index (i.e. assuming the oscillator strength to be zero). In this setting, the observed spectral shift of the cavity mode (in the weak coupling regime) gives a measure of the renormalization of the bare cavity mode in the presence of the active material. Our simulation yields, that a 0.8 nm thick layer of Wse$_2$ exactly reproduces the indicated observed energy jump.

The emergence of two distinct regions in our device with different coupling strength, modified detuning, and slightly changed exciton energies gives us the possibility to trace the polariton energy in our sample across the interface region between monolayer- and multilayer area (see sketch in Fig. 4a) via real-space resolved spectroscopy. In Fig. 4b, we plot the energy of the polariton modes as a function of real space coordinate (white points are extracted peak positions at the respective position with a threshold of 8000 Counts). As we trace the polariton energy from the heterostructure- to the single monolayer region, we clearly capture a step in the polaritonic energy at the interface with a height of 30 meV. This step height precisely matches the energy offset of the two polaritonic dispersions depicted in Fig 3c. i.e. it results from the local modification of the coupling strength V (following the dependence with the numbers of monolayers ) $V \propto n^{1/2}$, in conjunction with the change in detuning resulting from the increased cavity thickness stemming from the additional WSe$_2$ layer and different exciton positions in monolayer and the heterostructure region. The precise contributions to the spectral shift are summarized in Table 1. While the exciton shifts by $\sim 16$ meV (positive sign corresponds to lower energy), the cavity shifts $\approx 22$ meV at $k = 0\,\mu m^{-1}$. The detuning ($k = 0\,\mu m^{-1}$) for the monolayer and heterostructure is $\approx 11$ meV and $\approx 17$ meV, respectively. Therefore, the exciton and bare cavity energies yields a redshift of the lower polariton branch by $\approx 19$ meV for the heterostructure (Row 3), whereas the shift resulting from the increased coupling strength is $\approx 12$ meV (Row 5).

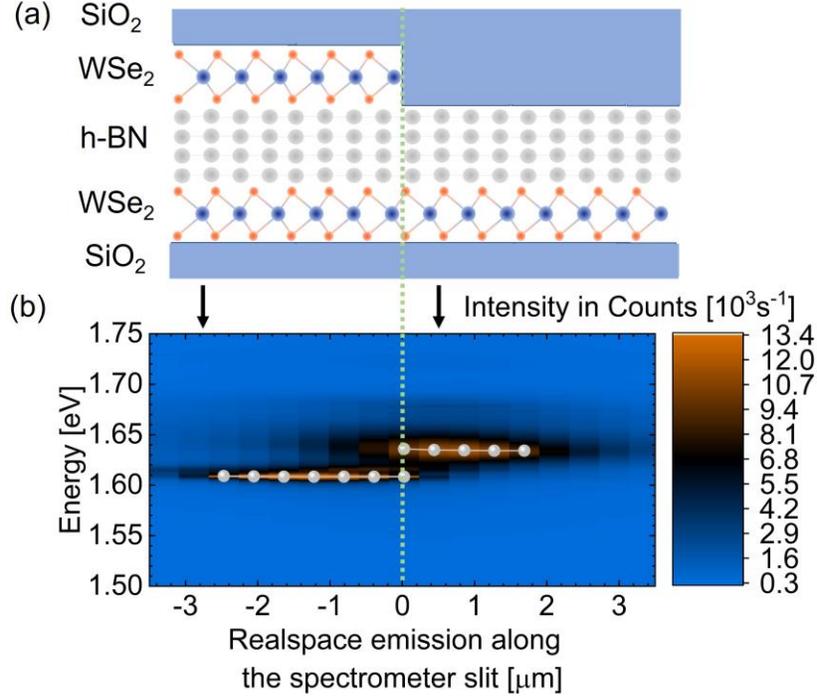

Fig. 4. a) Artistic drawing of the interface with varying coupling constant and detuning. b) Real-space resolved photoluminescence trace of the polariton resonance across the interface. Gray data points indicate position of maximum emission. The green dashed line indicates the interface between monolayer and heterostructure region and the black arrows connect the corresponding region with the photoluminescence spectrum.

|  | Monolayer | h-BN separated monolayers |
|---|---|---|
| Exciton position $E_{ex}$ / eV | 1.6722 | 1.6564 |
| Cavity position $E_C$ / eV | 1.6612 | 1.6396 |
| $0.5(E_{ex}+E_C)$ / eV | 1.6667 | 1.6480 |
| Detuning $\Delta = E_{ex} - E_C$ / meV | 11.0 | 16.8 |
| $0.5\sqrt{V^2+\Delta^2}$ /meV | 23.7 | 35.8 |
| $V/2$ / meV | 23.1 | 34.8 |

**Table 1. Spectral shift contributions of the potential step**

## 4. Conclusion

In conclusion, we have discussed the capability to engineer polaritonic potentials by aligning atomically thin crystals in a van-der-Waals heterostructure. At positions where the number of monolayers changes by $\pm 1$, the polariton energy experiences a shift that is composed of the interplay between change in detuning and coupling strength. We show, that this shift can be as large as 30 meV. We further note, that the characteristic Rabi-frequency also is digitally changed at the interface, and we believe that this could yield interesting phenomena in the dynamic behavior of polaritonic modes at the interface area. We also believe, that devices like the one demonstrated in this work can be utilized to investigate polariton dynamics, such as diffusion and expansion in atomically thin crystals, including measurements such as the polariton spin-torque in ballistic experiments, which should expose the interplay between polarization and band topology [24].

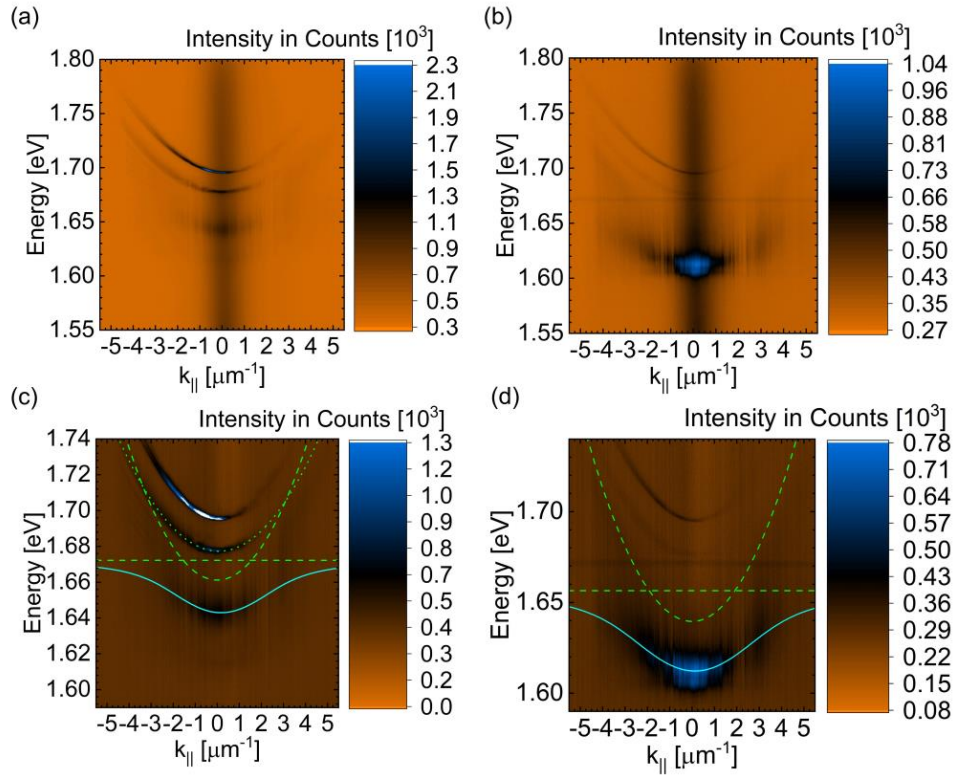

## Appendix A. Polariton dispersion – raw data

Fig. 5. A) and b) show the monolayer and separated layers polariton dispersion in momentum space. C) and d) show dispersions, where a uniform background over energy was subtracted from all linespectra. Blue lines correspond to the coupled oscillator fit and the green dashed lines show the exciton as well as the cavity resonance. The green dotted parabola corresponds to an empty cavity resonance with h-BN.

## Appendix B. Reflection contrast of the monolayer

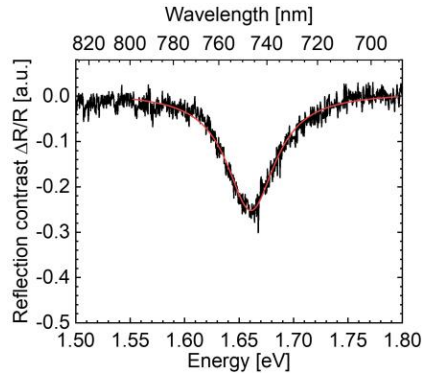

Fig. 6. Shows the reflection contrast of the bare monolayer at a slightly different position than in Fig.2a). While in Fig.2 the minima in a) and b) differ by ≈ 17 meV, the difference between

Fig. 6 and Fig. 2b) at the minima is ≈ 7 meV. Therefore inhomogeneities in the sample structure also influence the absorption shift.

**Funding**

C.S. acknowledges funding by the European Research Council within the Project Unlimit2D (Grant Agreement Number 679288); FE and HK acknowledge support by the Federal Ministry for Education and Research (BMBF) under Grant-ID 13XP5053A. S.T. acknowledges support by the NSF (DMR-1955668 and DMR-1838443). Core Research for Evolutional Science and Technology (JPMJCR15F3).

**Acknowledgements**

This work has been supported by the State of Bavaria. K.W and T.T. acknowledge support from the Elemental Strategy Initiative conducted by the MEXT, Japan

Correspondence and requests for materials should be addressed to Christian Schneider (christian.schneider@physik.uni-wuerzburg.de).

**Disclosures**

The authors declare no conflicts of interest.